\documentclass[twoside,12pt]{article}
\input epsf.tex
\usepackage{amsfonts,amssymb,amscd,amstext}
\usepackage{a4,theorem}
\fussy
\makeatletter
\@addtoreset{equation}{section}
\makeatother

\newcommand\be{\begin{equation}}
\newcommand\ee{\end{equation}}

{
\theoremstyle{break}
\theorembodyfont{\normalfont}
\theoremheaderfont{\scshape}
\newtheorem{theorem}{Theorem}[section]
\newtheorem{proposition}{Proposition}[section]
}

\newenvironment{proof}{\list{}{\setlength{\leftmargin}{0pt}}\item\relax
\noindent\emph{Proof.~}}{\endlist}

\begin{document}
\thispagestyle{empty}
\def\ee{\end{equation}}
\def\be{\begin{equation}}
\font\sevenrm=cmr7  
%
%
\def\IC{{\Bbb C}}
\def\IR{{\Bbb R}}
\def\IZ{{\Bbb Z}}
\def\IN{{\Bbb N}}
\def\IH{{\Bbb H}}
\def\II{{\Bbb I}}
\def\CG{{\cal G}}
\def\CH{{\cal H}}
\def\CA{{\CA}}
\def\CC{{\cal C}}
\def\CM{{\cal M}}
\def\CA{{\cal A}}
\def\CG{{\cal G}}
\def\CE{{\cal E}}
\def\CN{{\cal N}}
\def\Cm{{\goth m}}
\def\Dsl{\,\raise.15ex\hbox{/}\mkern-13.5mu D}
\def\dsl{\raise.15ex\hbox{/}\kern-.57em\partial}
\def\qft{{quantum field theory}}
\def\Qft{{Quantum Field Theory}}
\def\h{{\goth h}}
\def\Tr{\hbox{{\rm Tr}}}
\def\tr{\hbox{{\rm tr}}}
\def\Ker{\hbox{{\rm Ker}}}
\def\End{\hbox{{\rm End\,}}}
\def\ad{\hbox{{\rm ad\,}}}
\def\ha{{1\over2}}
\def\circle{\circ}
\def\mapright#1{\smash{\mathop{\longrightarrow}\limits^{#1}}}
\def\half{{\textstyle{1\over2}}} 
\pagestyle{myheadings}
%
\centerline{\Large {Vacuum Nodes and Anomalies in Quantum Theories
\footnote{Fax: 34 976 761264, Email:asorey@saturno.unizar.es}}}
\bigskip
\centerline{ {\sc {M. Aguado, M. Asorey and J.G. Esteve}}}

\begin{center}{\it Departamento de F\'{\i}sica Te\'orica.
 Facultad de Ciencias\\
Universidad de Zaragoza.
50009 Zaragoza. Spain} \end{center}

\setcounter{page}{1}

\begin{abstract}
We show that nodal points of ground states of some
quantum systems with  magnetic interactions 
can be identified in  simple geometric terms.
We analyse in detail two different archetypical systems:
i) the planar rotor with a non-trivial magnetic flux  $\Phi$
and ii) the Hall effect on a torus.
In the case of the planar rotor we show that the level repulsion 
generated by any reflection invariant potential $V$ is encoded in 
the nodal structure of the unique vacuum for $\theta=\pi$. In the 
second case we prove that the nodes of the first Landau level for unit 
magnetic charge appear at the crossing of the two non-contractible
circles $\alpha_-$, $\beta_-$ with holonomies $h_{\alpha_-}(A)=
h_{\beta_-}(A)=-1$ for any reflection invariant potential $V$. This 
property illustrates the geometric origin of the quantum translation 
anomaly.

\end{abstract}

\section{Introduction}
Classical configurations play different 
roles in the description of quantum effects. Monopoles, skyrmions, vortices, 
solitons, kinks and  similar classical configurations contribute to
unveil the existence of non-trivial sectors in the energy spectrum
of many quantum theories. Classical configurations are also important for the description
of superselection sectors and non-trivial phase structures in quantum
field theories. Tunnel effect is semiclassically described by means of instantons.  There is, however, another kind of classical configurations which
play a genuine quantum role in the description of some physical effects:
the nodal configurations of physical states. It is known that the structure 
of those nodes encode information about relevant physical properties such as 
the complete integrability or chaotic behaviour of the corresponding systems. 

Standard minimum principle arguments disfavor the appearance of nodes
in ground states of quantum systems.
However, the presence of CP violating interactions
invalidates the use of such arguments  and the vacuum
response to this kind of interactions may involve the
appearance of nodes. 
The existence  of a non-trivial nodal structure in the
vacuum states of quantum theories with CP violating interactions 
provides a new perspective in the analysis of the role of
classical configurations in the quantum theory. 
In some cases, the infrared behaviour of
the theory is so dramatically modified by the CP violating
interaction that a confining vacuum state can become non-confining.
The connection between the absence of confinement
and the existence of nodes in the vacuum state, suggests that
new classical field configurations related to the nodal
structure of the quantum vacuum emerge as  new candidates to play a 
significant role in the mechanism of confinement. This idea has been
successfully exploited to show the absence of spontaneous breaking
of CP symmetry at $\theta=\pi$ for various field theories
\cite{9} \cite{10}. 

In this paper we analyse the connection  between
the appearence of nodes
in ground states of quantum systems generated by CP violating interactions
and some non-perturbative quantum effects. In particular,
we analyse in some detail the vacuum nodal structure of the quantum
planar rotor with a $\theta=\pi$ term and the quantum Hall effect on a torus. In both cases the vacuum nodal structure turns to be
intimately related to the behaviour of the corresponding ground states under CP symmetry or translation symmetry.

In order to understand the physical origin of vacuum nodes
let us briefly recall the standard argument which prevents the vanishing of the vacuum.

Let us consider a quantum system evolving on a finite
dimensional Riemannian manifold ($M,g$) with Hamiltonian 
\be 
H={1\over 2 } \Delta + V(x),
\label{ 1.1}
\ee
defined by the Laplace-Beltrami operator $\Delta$ and a non-singular
potential $V(x)$. Unitarity of  quantum evolution requires 
 the potential $V(x)$ to be real, $V(x)=V(x)^\ast$, to guarantee
the hermicity of $H$. In this case the system is invariant under time reversal,
		$$U(T)\psi(x,t)=\psi(x,-t)^\ast,$$
because $[U(T),H]=0$. This symmetry implies that for any energy level
there is a basis of real stationary states, $\psi_n(x)=\psi_n(x)^\ast$.
 Indeed, if $\psi_n(x)$ is
an eigenstate of $H$ with energy $E_n$, the state $\psi_n(x)^\ast$ is also
an eigenstate with the same energy. If $\psi_n$
is not real, $\psi_n^\ast\neq  \psi_n$, the states
$\psi_\pm=\psi_n^\ast\pm  \psi_n$ will have the same
energy $E_n$ and will be real, irrespectively of the  degeneracy or
not of the energy level. 

If $H$ is bounded
below and  has a non-trivial  discrete
spectrum there is a ground state $\psi_0$ whose energy attains the minimum
$E_0$ of the energy spectrum. Because $V$ has no singularities on $M$
it is trivial to see that  $\psi_0$ cannot vanish for any point $x$
of the configuration space. Indeed, $\psi_0$ satisfies the stationary
equation
$$H\psi_0=E_0\psi_0.$$ If the set  of nodal points  of $\psi_0$,
$\CN_0=\{x\in M; \psi_0(x)=0\}$, is non-trivial, the
positive real function $\psi_1=|\psi_0|$ is smooth everywhere
except at the points of $\CN_0$. The expectation value of the
Hamiltonian on the state $\psi_1$ is again $E_0$ because the 
delta function singularity of $\Delta\psi_1$ at 
$\CN_0$ is cancelled by the vanishing of $\psi_1$ at that point,
$$E_0= <\psi_0|H \psi_0>= <\psi_1|H \psi_1>;$$
e.g. in one dimension, if there is a nodal point, $\CN_0=\{x_\ast\}$, we have 
 $$ <\psi_0|{d^2\over dx^2} \psi_0>= <\psi_1|{d^2\over dx^2} \psi_1>
-2 \int^\infty_{-\infty}\psi_0(x)^\ast \delta(x-x_\ast) {d\over dx} 
\psi_0(x)=<\psi_1|{d^2\over dx^2} \psi_1>.$$
Since $E_0$ is the lowest eigenvalue of $H$ this
means that $\psi_1$ is also an eigenstate of $H$ with the same
energy that $\psi_0$.
Now, elliptic
regularity implies that any eigenstate of $H$ must be smooth, thus
$\psi_1$ cannot be an eigenstate because its  differential is
discontinuous at $\CN_0$. The contradiction, being motivated by
the assumption of existence of nodes, disappears if $\psi(x)\neq 0$
for any $x\in M$. The same argument leads to the proof of vacuum
uniqueness. If the vacuum were degenerate, there  would exist
another ground state $\psi_1\neq \psi_0$. Then, the ground state
defined by $\chi(x)=\psi_0(x_\ast)\psi_1(x)-\psi_1(x_\ast)\psi_0(x)$
will vanish for $x=x_\ast$, which cannot occur by the previous
argument. 

Both results  rely heavily on the real, local and smooth
characteristics of the potential $V$. Exceptions for this archetypical
infrared behaviour of  quantum systems can arise either by the
introduction of internal degrees of freedom (e.g. spin), singular or
non-local potentials, or complex interactions. 

Complex interactions are physically generated by
 the presence of magnetic fields. The interaction
with the magnetic gauge field potential $A$ through the gauge
principle  of minimal coupling leads to a Hamiltonian
\be H_A=
{1\over 2 } \Delta_A + V(x),
\label{ 1.2}
\ee
which is not invariant under time reversal, $U(T) H_A U(T)=H_{-A}$.
The eigenstates are not necessarily real functions and the rest of
the argument leading to the absence of nodes and uniqueness of the
vacuum state fails.

\section{The Planar Rotor}

Let us consider the case of a charged particle moving on a circle under
the action of a periodic potential $V(\varphi)$ and a non-trivial
magnetic flux $\Phi=2\pi\epsilon$ crossing through the circle. In
this case, $M=S^1$ and
\be H_\epsilon=-{1\over 2 } (\partial_\varphi-i\epsilon)^2 +
V(\varphi), \label{ 2.1}\ee
where $\varphi\in [-\pi,\pi)$ is the angular coordinate  of the
circle, and we assume that the mass and charge of the particle are
$m=e=1$. 

\begin{proposition}
If the potential $V$ is reflection invariant $V(\varphi)=V(-\varphi)$,
the matrix element 
$$K^\epsilon_{\hbox{\sevenrm{T}}}(\varphi_0,\varphi_1)=\hfill\break
<\varphi_0|{\textstyle
e}^{-TH_\epsilon}| \varphi_1>$$
of the heat kernel operator 
vanishes for $\epsilon=1/2$ when $\varphi_0=0$ and $\varphi_1=\pi$,
i.e.  $$K^{^{1/2}}_{\hbox{\sevenrm{T}}}(0,\pi)=0.$$ 
\end{proposition}

\begin{proof}
In such a case the Hamiltonian (2.1)
is invariant under the Bragg reflection symmetry 
$$U(P)\psi(\varphi)=\psi(-\varphi)$$ 
and  it
is always possible to find in the  Hilbert space $\CH=L^2(S^1)$ a
complete basis of stationary states with definite U(P) symmetry. If
the energy level is not degenerate,  the corresponding physical
state $\psi(\varphi)$ has to be either  even or  odd under $U(P)$
symmetry. In
the degenerate case, if $U(P)\psi$ is not the same state that
$\psi$, the  stationary functionals $\psi_\pm= \psi\pm U(P)\psi$ are
parity even/odd, respectively. This implies that kernel  element
$K^{^{1/2}}_{\hbox{\sevenrm{T}}} (0,\pi)$ is reflection invariant
\be
U(P)^\dagger K^{^{1/2}}_{\hbox{\sevenrm{T}}} U(P)(0,\pi)=
\sum_n U(P) \psi_n(0)^\ast U(P)\psi_n(\pi) {\hbox{\rm
e}}^{-E_n T}=K^{^{1/2}}_{\hbox{\sevenrm{T}}} (0,\pi),
\label{ 2.2}
\ee
\noindent
 On the other hand in the path
integral representation 
\be
K^\epsilon_{\hbox{\sevenrm{T}}}(\varphi_0,\varphi_1)= \int_{{\phantom{LLLLL}\atop{ \varphi(0)=\varphi_0\atop 
{\varphi(T)=\varphi_1}}}}
 \!\!\!\!\!\!\!\!\!\!\!\!\!\!\!\!\!
\delta \varphi\ \ \ \ \ {\hbox{\rm exp}}\left\{ -{\int_0^T dt 
\Bigl[\half\dot\varphi(t)^2 +i\epsilon \dot\varphi(t)
+V(\varphi(t))\Bigr]}\right\}. \label{ 2.3}
\ee
we have that
\be
U(P)^\dagger K^\epsilon_{\hbox{\sevenrm{T}}} U(P)(0,\pi)=
K^{-\epsilon}_{\hbox{\sevenrm{T}}}(0,\pi)
\label{ 2.4}
\ee
because 
the P transformation leaves the points $\varphi= 0$ and $\varphi=
\pi$ invariant but changes the sign of the $\epsilon$--term
 in the exponent of the  path integral, since it reverses the
orientation of every path. The contribution of this term
 becomes $-2\pi i\epsilon({{1/2}}+n)$
instead of $2\pi i\epsilon({1/2}+n)$ for any trajectory  $\varphi(t)$
in $S^1$ connecting $\varphi=0$ with  $\varphi=\pi$ with winding
number $n$. Thus,  the kernel  element 
$K^{^{\epsilon}}_{\hbox{\sevenrm{T}}}(0,\pi)$ is not invariant under reflection
symmetry, unless  $\epsilon=0$ (mod.
$\IZ$).  In particular  for $\epsilon={1/2}$, the
kernel element $K^{^{1/2}}_{\hbox{\sevenrm{T}}}(0,\pi)$  is parity odd and  purely imaginary
$$
U(P)^\dagger K^{^{1/2}}_{\hbox{\sevenrm{T}}} U(P)(0,\pi)=
K^{-{^{1/2}}}_{\hbox{\sevenrm{T}}}(0,\pi)=K^{{^{1/2}}}_{\hbox{\sevenrm{T}}}(0,\pi)^\ast=
-K^{{^{1/2}}}_{\hbox{\sevenrm{T}}}(0,\pi). $$
This is in disagreement with (2.2) unless the kernel  element vanishes
for those points  \hfill\break $K^{^{1/2}}_{\hbox{\sevenrm{T}}} (0,\pi)=0.$ 
\end{proof}

This property is
independent of the potential term $V$ and the value of $T$.
In particular, it implies that the same vanishing property
holds for the restriction of $K_{\hbox{\sevenrm{T}}}
(0,\pi)$ to any energy level, e.g
the ground state. If the vacuum is non degenerate it has to
vanish  either at $\varphi=\pi$ or $\varphi=0$ for this
particular value $\Phi=\pi$ of the magnetic flux
($\epsilon=1/2$).

This property of the heat kernel can also 
be understood in the Hamiltonian formalism.
The presence of the magnetic flux has a non-trivial effect
in the energy spectrum of the theory
(Aharonov-Bohm effect) because of the non-simply connected character
of $S^1$, $\pi_1(S^1)=\IZ$. Although the $\epsilon$ dependence 
cannot be removed by a globally defined gauge transformation, the
singular gauge transformation
\be
{\xi(\varphi)={\hbox{\rm e}}^{-{i\epsilon}\varphi}  \psi(\varphi)}
\label{ 2.5}
\ee
which is uniquely defined on the domain $(-\pi,\pi)$ but 
is
discontinuous at $\varphi=\pm \pi$, removes the $\epsilon$ dependence of 
quantum Hamiltonian
$$
{\widetilde{H}_\epsilon= 
{\hbox{\rm e}}^{-{i\epsilon}\varphi}
H_\epsilon {\hbox{\rm e}}^{{i\epsilon}\varphi}=H_0}.$$
 The $\epsilon$
dependence is, however, encoded in the non-trivial boundary
conditions that physical states have to verify  at the boundary
$\varphi_\pm=\pm\pi$,
\be
{\xi(-\pi)={\hbox{\rm e}}^{-{i\epsilon}} \xi(\pi).}
\label{ 2.6}
\ee
In this sense the transformation  is trading the 
$\epsilon$--dependence of the Hamiltonian for non-trivial boundary 
conditions at $\varphi_\pm$.

The relevant extra property which  allows us to extract some
information on the  nodal structure of the quantum vacua is that
the  theory is U(P) invariant for $\epsilon=1/2$. In the Hamiltonian
approach this property of the special case $\epsilon=1/2$ comes from
the fact that the boundary condition (2.6), becomes an
anti-periodic boundary condition, 
${\xi(\varphi_+)=- \xi(\varphi_-),}$
 which is a reflection invariant condition. 

 As discussed above it is 
always possible to find a 
complete basis of stationary states with definite U(P) symmetry. If
the energy level is not degenerate the corresponding physical
state $\psi(\varphi)$ has to be U(P) even or U(P) odd. In the
degenerate case, we can have states with both parities.
But,
because of  anti-periodic  boundary conditions, any of them
satisfies that
$
U(P)\xi( \pi)= \xi(-\pi)
 =-\xi(\pi).
$
Thus, for any  parity even state $\xi_+$ this is 
possible only if $\xi_+$
vanishes for $\varphi=\pm \pi$, $\xi_+(\pm \pi)=0$.
In the same way since for any parity odd state $\xi_-$ we have 
$
U(P)\xi_-(0_-)= \xi_-(0_+)= \xi_-(0),
$
any parity odd state vanishes for $\varphi=0$,i.e. $\xi_-(0)=0$.
This property explains the vanishing of the heat kernel element
$K^{\hbox{\sevenrm{1/2}}}_{\hbox{\sevenrm{T}}}(0,\pi)$ 
for any value of $T$, previously derived by path integral methods,
because {\it half} of the states of an orthonormal basis of 
stationary states in $L^2(S^1)$ vanish at
$\varphi=0$ whereas the other {\it half} 
vanish at $\varphi=\pi$.

Let us now consider the structure of the 
ground state $\psi_0$.

\begin{proposition}
A planar rotor interacting with a transverse magnetic flux $\Phi=\pi$
and a reflection invariant non-constant potential $V(\varphi)$ with maximum
height at  $\varphi=\pi$ and minimum value  at $\varphi=0$ has a
unique vacuum state $\psi_0$ which is parity even and vanishes at
$\varphi=\pi$.  
\end{proposition}

\begin{proof}
Since the potential term $V$ is non-trivial it 
gives a non-trivial  contribution to the energy of stationary states.
 The states with lowest energy which are parity even vanish at
$\varphi=\pm \pi$, where the potential terms attains its maximal
value, and  cannot have the same energy as parity odd states which vanish
at  $\varphi=0$, where the
potential terms attains its minimal value. This feature implies
that the quantum vacuum state  $\psi_0$ is non degenerate, is
parity even and  vanishes at $\varphi=\pm\pi$. The splitting of
energies between the ground state and the first excited state
can also be understood in terms of tunnelling effect induced by
instantons. But the argument used above is completely rigourous
and does not rely on any semiclassical approximation or
asymptotic expansion (see \cite{6} for an early 
anticipation of this behaviour of the ground state based in 
numerical calculations).
\end{proof}

 The existence of a non-trivial potential with such a peculiar behaviour 
 is  crucial for the proof of the existence of vacuum nodes.
If $V=0$, there is no splitting between the energies of even and odd
states and the ground state becomes degenerate. In this case there
are ground states with indefinite parity which are linear
combinations of parity even and parity odd ground states and have no
nodes. However, the kernel of the restriction of the operator 
$K^{^{1/2}}_{\hbox{\sevenrm{T}}}$ to
the ground state subspace also vanishes for the pair of points $0$ and $\pi$ 
as the path integral formula predicts. The parity of the vacuum for
general potentials with unique vacuum, depends on the structure of
the potential. Although small generic perturbations of potentials of the
type considered in Proposition 2.2 preserve the even character of
the vacuum it might change for large perturbations due to the appearence
of level crossings. The existence of such crossings for $V=0$ guarantees
the consistence of the result when the maximum and minimum of $V$ are
interchanged.

The presence of a magnetic field generates nodes in
the ground state as a vacuum response to the magnetic flux crossing
the circle where the system evolves. This system mimics the behaviour
 of the
1+1 dimensional QED on a cylinder with a $\theta$--term
$\theta=2\pi\epsilon$ when $V=0$.

\section{ Hall Effect in a Torus}

A charged quantum particle ($e=m=1$) moving on a two-dimensional torus
$T^2$ under the action of an uniform magnetic field $B= k/2\pi$
($k\in \IZ$) and an external potential $V$ is governed by the  Hamiltonian 
\be
{{{H_A}}= -{1 \over 2}  \Delta_A} + V,
\label{ 3.1}
\ee
where $\Delta_A$ is the covariant Laplacian with respect to a
U(1) gauge field $A$ with curvature $B$ defined 
on the line bundle $E_k(T^2,
\IC)$
 with first Chern class $k\in\IZ$, whose sections are the
quantum states. 

For trivial potentials $V={\rm const}$, the spectrum of the Hamiltonian 
(3.1) is exactly solvable. It is given by the Landau
levels 
$$E_n = {B} \left(n+ {1 \over 2}\right) \qquad n \in \IN$$
as in the infinite plane case. However, in the present case the
degeneracy of each level is finite, $\dim {\CH_0}=\vert k \vert$, whereas  
in infinite volume the degeneracy is infinite for $k\neq 0$.
	The degeneracy of the ground state $E_0$ is not only dependent  
of the first Chern class of the line bundle  $E_k(T^2, \IC)$, where physical states are defined but also of the background metric of the torus 
and the form of the magnetic field \cite{3}.

\subsection {Translation anomalies}

In the presence of the constant magnetic field, $V={\rm const}$, and a 
symmetric metric
the classical
system is translation invariant but the 
quantum generators of translation
symmetries given by 
$${\hbox{\bf{L}}}^j= -i D^j_{A}-\epsilon^{jl}x_l B\qquad j=1,2$$
suffer from an anomaly which transforms the abelian algebra
$\IR \times \IR$ into the  Heisenberg
algebra 
\be
{\lbrack {\hbox{\bf{L}}}_1, {\hbox{\bf{L}}}_2 \rbrack = -i B}
\label{ 3.2}
\ee
as  a central extension  with  central charge $B$. This is easy to
understand because the system has two degrees of freedom and cannot
have three independent conmuting operators corresponding to
time and space translations (3.1) (3.2).

In a $T^2$ torus there is an extra anomaly of translation symmetry, for
 if the Heisenberg algebra were a real symmetry of the
quantum  system  the energy 
levels would be infinitely degenerate, since 
any energy level supports a representation of the symmetry algebra
and any representation of the Heisenberg algebra  (3.1)  must be
infinite dimensional, but the  energy levels do have a 
finite degeneracy $k$.

The presence of the anomaly is explicitely shown by the existence
of a non-vanishing correlation function involving the time
derivative of the would be conserved currents $l_j = \dot{x}_j -  
\epsilon_{jn}x^n B$ associated to translation
transformations
$$<\psi_0|\dot{l}_j(t) \dot{x}_n(s)| \psi_0 > =
 -i \theta_3(0)(\delta_{jn}+i\epsilon_{jn}) {B\over 2}^2{\rm e}^{iB (s-t)}, 
 $$
where 
$$\theta_3(u)= \sum_{n=-\infty}^\infty {\rm e}^{-\pi n^2/2} 
{\rm e}^{2nui}$$
is the third Jacobi theta function. This is the simplest example
of an anomalous   symmetry in a quantum mechanical system.
Notice that it is not present in the infinite volume limit.
 
There is an operator theory explanation for this anomaly \cite{4}
\cite{1}.
 Although the
generators of the translation Heisenberg algebra ${\hbox{\bf{L}}}_1,
{\hbox{\bf{L}}}_2$ commute with ${\hbox{\bf{H}}}$  on the domain 
of functions with compact 
support on $(0, 2 \pi) \times  (0, 2 \pi)$, the corresponding
selfadjoint extensions  {\it {do not}} commute, because the domain of
definition of ${\hbox{\bf{H}}}$ is not preserved by the action
of ${\hbox{\bf{L}}}_j$. In  this sense translation invariance is
broken in the quantum system.
This interpretation of the anomaly based on the anomalous 
behaviour of
the domain of definition of the quantum Hamiltonian under
translations was first pointed out by Esteve \cite{15} and
Manton \cite{21}.
In this case the existence of an anomalous commutator 
is crucial for the understanding of
the finite degeneracy of energy levels in spite of 
the existence of a partial translation invariance.

There is a simpler geometrical interpretation of the anomaly.
The quantum system is not completely specified  by the magnetic
flux. To define the connection $A$ one has to specify the
holonomies, $h_\alpha(A)$, $h_\beta(A)$,
along two complementary non--contractible circles of the torus
$\alpha$, $\beta$. Once  $h_\alpha(A)$,
$h_\beta(A)$ are specified, the holonomy along any other closed
loop on $T^2$ is completely determined because the  holonomies along
two homotopically equivalent circles differ by a phase factor whose
exponent is twice  the magnetic flux of the torus domain enclosed by
them. This means  that while any of the basic circles
$\alpha$, $\beta$ sweeps the torus under a $2\pi$ translation its
holonomy describes a non-contractible 
loop along  the gauge group $U(1)$.
Thus, there are at least two non--contractible circles $\alpha_0$,
$\beta_0$ on the torus at which the holonomies of $A$ reduce to the
identity, i.e. $h_{\alpha_+}(A)=h_{\beta_+}(A)=I$.  In a similar way
there are at least other two non-contractible circles $\alpha_-$,
$\beta_-$ at which the  the holonomies of $A$ are minus the identity,
i.e. $h_{\alpha_-}(A)=h_{\beta_-}(A)=-I$ (see fig. 1). 

 \vskip 0.2cm
{\hskip2.8cm\epsfbox{sin2.ai}}
\vskip 0.1cm
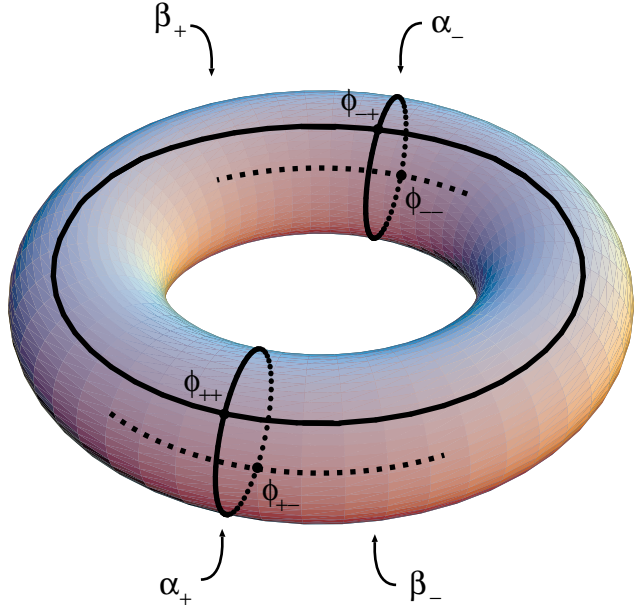
\centerline{\hbox{\rm {\bf Figure 1.} Circles with holonomies $\pm 1$
and nodal points of quantum vacua}}
\vskip 0.3cm

 These loops are unique for any connection with unit first Chern class, i.e. $c(A)=k=1$. The existence of such special loops
explains why translation symmetry is completely broken for $k=1$
in the Hall effect on a torus. Only translations which give a
complete turn to the torus leave the Hamiltonian invariant. The
translation symmetry group  is  then reduced  from $T^2$ to
$I$. For the same reasons for higher values of $k$ the number of
closed circles with trivial holonomy in each homotopy class is equal
to $k$. If $k>1$ there are $k$  circles in the same homotopy
class with the same holonomy. This means that the continuous
translation symmetry is reduced by the anomalies to a discrete
quantum symmetry generated by translations by an angle
$2\pi/k$ in each of the two transversal directions of the torus, i.e.
the symmetry is reduced from  $T^2$ to a central extension of $\IZ_{k-1}\times
\IZ_{k-1}$.

\subsection {  Parity anomaly}

There is another discrete symmetry which also becomes anomalous upon
quantization. Let us introduce angular coordinates on the torus,
$T^2=[0, 2\pi)\times [0, 2\pi)$
The classical system is invariant under the combined action of 
two reflections with respect to any pair of angles
$\phi=(\phi_1,\phi_2)$,
$$
P_{\phi_1}(\varphi_1,\varphi_2)=(2\phi_1-\varphi_1,\varphi_2)\qquad
P_{\phi_2}(\varphi_1,\varphi_2)=(\varphi_1, 2\phi_2-\varphi_2)$$
i.e. $P_{\phi}=P_{\phi_1} P_{\phi_2}$. 
Notice that any of these parity transformations $P_{\phi_1},
 P_{\phi_2}$ is not a
symmetry because it reverses the orientation of the torus and, thus,
the sign of the magnetic field.

However, the need of specification in the quantum theory of
the holonomies $h_\alpha(A)$, $h_\beta(A)$ breaks down this 
 reflection symmetry  with respect to 
the a generic point of the torus $(\phi_1, {\phi_2})$
except for the four crossing points  
 $\phi_{++}=(\phi_1^+,\phi_2^+),
\phi_{+-}=(\phi_1^+,\phi_2^-)$,
$\phi_{-+}=(\phi_1^-,\phi_2^+), \hfill\break
\phi_{--}=(\phi_1^-,\phi_2^-)$ of the  circles with
holonomies $I$ or $-I$ (see Fig. 1), 
$$\alpha_\pm(\varphi)= (\phi^\pm_1,\varphi)\qquad
 \beta_\pm(\varphi)= (\varphi,\phi^\pm_2)
 \qquad \varphi\in [0,2\pi). $$

Reflection with respect to any other
point  transforms loops  into  loops with different
holonomy for $k=1$.  For $k>1$ there are more crossing points of
circles with holonomies $I$ and $-I$, thus, the reflection
symmetry group is bigger in that case. In any case the remaining
quantum symmetry, $U(P_{\pm\pm})$ defined by 
\be
U(P_{\pm\pm})\psi(\varphi_1,\varphi_2)=
{\hbox{\rm
e}}^{-i(\phi_2^\pm\varphi_1-\phi_1^\pm\varphi_2)}\psi
\left(U(P_{\pm\pm})(\varphi_1,\varphi_2)\right) 
\label{ 3.3}
\ee
is very relevant to find the nodes of
the ground states.

\subsection{Vacuum Structure}

Since the line bundle $E_k(T^2, \IC)$ is non-trivial for
$k\neq 0$ any section must have nodal points. This means that
any energy level has a non-trivial nodal structure. For $|k|>1$
the degeneracy of energy levels is $k$ which means that given
any point $\phi_\ast$ on the torus there is one state in that  level
with a node at $\phi_\ast$. Therefore in such a case the physical
meaning of the nodal configuration cannot be relevant. However,
in the case $k=1$ ($B=1/2\pi$) there is no degeneracy in the energy
levels and the vacuum  do have only one node which certainly has to
be a very distinguished classical configuration for the quantum
system. The search of vacuum nodes is simplified for the case
of $P_{++}$-symmetric potentials $V(P_{++}\varphi)=V(\varphi)$, 
by the following result.

\begin{proposition}
For any $P_{++}$ symmetric potential $V$
the heat kernel elements
\be
K^A_{\hbox{\sevenrm{T}}}(\phi_{++},\phi_{--})=\\
K^A_{\hbox{\sevenrm{T}}}(\phi_{+-},\phi_{--})=\\
K^A_{\hbox{\sevenrm{T}}}(\phi_{-+},\phi_{--})= 0\\
\ee
vanish for any $T$ if $k=1$. 
\end{proposition}

\begin{proof} 
The basic strategy is similar to that used in the planar rotor.
Because of the  non-degeneracy of the energy levels, any
stationary state must have a definite $U(P_{\pm\pm})$-parity  
symmetry with respect
to the four points $\phi_{++},\phi_{+-},\phi_{-+},\phi_{--}$ where
the circles with holonomies $I$ and $-I$ cross each other. There are
four quantum parity symmetry transformations $U(P_{{++}})$,
$U(P_{+-})$, $U(P_{-+})$ and
$U(P_{--})$. Although,  the four transformation are
identical in $T^2$, e.g. all of them leave the four points invariant, they
define four different unitary transformations in the space of quantum
states. If we redefine our coordinates so that
$\phi_1^+=\phi_2^+=0$, we have that  $\phi_1^-=\phi_2^-=
\pi$. In such coordinates $\varphi=(\varphi_1,\varphi_2)$ the gauge
field with the required holonomy properties is given by   
\be A_i={B\over
2}\epsilon_{ij}\varphi^j 
\label{ 3.4}
\ee
in a gauge with boundary
conditions
\be \psi(\varphi_1+2\pi,\varphi_2)={\hbox{\rm e}}^{i\pi
B\varphi_2}\psi(\varphi_1,\varphi_2)\qquad 
\psi(\varphi_1, \varphi_2+2\pi)={\hbox{\rm e}}^{-i\pi
B\varphi_1}\psi(\varphi_1,\varphi_2).
\label{ 3.5}
\ee
Since the $P_{++}$ symmetry leaves the point $\phi_{--}$ 
invariant, physical states $\psi$ must verify 
$U(P_{++})\psi(\pi,\pi)=\psi(-\pi,-\pi)= {\hbox{\rm e}}^{i
\pi}\psi(\pi,\pi)=-\psi(\pi,\pi).$
Thus, if $\psi$ is $U(P_{++})$ even $\psi$ has a node
at $\phi_{--}$, i.e. $\psi(\phi_{--})=0$.
For the same reason  
$$U(P_{{++}})\psi(0,0)
=\psi(0,0),\qquad U(P_{++})\psi(0,\pi)
=\psi(0,-\pi)= \psi(0,\pi),$$
$$U(P_{++})\psi(\pi,0)
=\psi(-\pi,0) =\psi(\pi,0),$$
 which implies
that  $P_{++}$ odd states must vanish at 
$\phi_{++}$, $\phi_{-+}$ and $\phi_{+-}$, i.e. $\psi(\phi_{++})=
\psi(\phi_{+-})=\psi(\phi_{-+})=0$.
In a similar way we get that
$$U(P_{--})\psi(\pi,\pi)=\psi(\pi,\pi),\quad 
U(P_{--})\psi(0,\pi)
={\hbox{\rm e}}^{i
\pi/2} \psi(2\pi,\pi)=-\psi(0,\pi),$$
$$
U(P_{{--}})\psi(0,0)=\psi(2\pi,2\pi)={\hbox{\rm e}}^{-i
\pi} \psi(0,2\pi)
=-\psi(0,0),
$$
$$
U(P_{--})\psi(\pi,0)={\hbox{\rm e}}^{-i
\pi/2} \psi(\pi,2\pi)=-\psi(\pi,0),$$ which implies
that  $P_{--}$ odd states must vanish at $\phi_{--}$, whereas
$P_{--}$ even states must vanish at
$\phi_{++}$, $\phi_{-+}$ and $\phi_{+-}$. Similar properties hold
for the remaining  parity operators  $U(P_{-+}),U(P_{+-})$.
Since there is a complete basis of stationary states consisting of
$U(P_{--})$ even and $U(P_{--})$ odd states, this implies the vanishing 
of the kernel matrix elements (3.4).

As in the planar rotor case there is an alternative derivation of 
 the same results. It is based on
the path integral approach. The
method also carries enough information  to identify
the parity of the vacuum state. The essential feature is
to prove that the  matrix element \hfill\break
$K^A_{\hbox{\sevenrm{T}}}(\phi_{++},\phi_{--})=<\phi_{++}|{\textstyle e}^{-TH_A}|
\phi_{--}>$ of the euclidean time evolution  kernel vanishes 
for any $T$. This property can be easily
derived from the path integral representation of the heat kernel
\be
K^A_{\hbox{\sevenrm{T}}}(\phi_{++},\phi_{--})
= \int_{{\phantom{LLLLL}\atop{ \varphi(0)=\phi_{++}\atop 
{\varphi(T)=\phi_{--}}}}}
 \!\!\!\!\!\!\!\!\!\!\!\!\!\!\!\!\!
\delta \varphi\ \ \ h^A_\varphi(t) \ \ {\hbox{\rm exp}}\left\{ -{\int_0^T dt 
\Bigl[\half\dot\varphi(t)^2  + V(\varphi(t))
\Bigr]}\right\}. \label{ 3.6}
\ee
where  $h^A_\varphi(t)$ is the holonomy of the closed path
$\varphi(t)$.
In the path
integral representation (3.7) a path $\varphi(t)$ connecting
$\phi_{++}$ and $\phi_{--}$ transforms under the $P$ reflection
symmetry into another path $P\varphi(t)$ which connects the same
points and  gives the same contribution to the
real term of the exponent in the path integral. However, the
contribution of both paths to the imaginary part  is different. They
contribute to the path integral with a phase factor which is exactly
the holonomy of $A$ along the paths. It is immediate to see that
the ratio of both contributions 
$h^A_\varphi(t) (h^A_{P\varphi(t)})^{-1}$ equals the holonomy of the
closed loop obtained by composition
$\varphi(t)\circle P\varphi(T-t)$ which is
 in the homotopy class $(2n_1+1,2n_2+1)$ of
$\alpha^{2n_1+1}_-\circle\beta^{2n_2+1}_+$. 
The holonomy  splits, by Stokes
theorem, into a factor which is the holonomy of
 the basic circles $\alpha^{2n_1+1}_-\circle \beta^{2n_2+1}_+ $ and the magnetic flux
$\Phi_1$, $\Phi_2$  crossing two surfaces $\CC_1$ and  $\CC_2$ in $T^2$: 
$\CC_1$ being the domain of $T^2$ enclosed by the curves 
$\varphi(t)$ 
and
\be
\gamma_1(t)=\left\{\begin{array} {ll} 
\displaystyle
((2 n_1+1)\pi+ (2 n_1+1),0)
{2\pi\over T} (t-{T\over 2}))& 
 {0\leq t \leq {T\over 2}} \\
 \displaystyle
 ((2 n_1+1)\pi,(2 n_2+1)\pi+ (2 n_2+1)
{2\pi\over T} (t-T)) &
 {T\over 2}\leq t\leq {T},\\ 
\end{array} \right.
\ee
and $\CC_2$ being the surface
 enclosed by $P\varphi(T-t)$ and $P\gamma_1(T-t)$ (see Fig. 2).
Those contributions of  magnetic fluxes are opposite and cancel
each other. Thus, the  contribution of $h^A_\varphi(t)
(h^A_{P\varphi(t)})^{-1}$ is reduced to the
holonomy of $\alpha^{2n_1+1}_-\circle\beta^{2n_2+1}_+$, i.e. 
$(-1)^{2n_1+1}=-1$. This means that the contributions of 
$\varphi(t)$ and $P\varphi(t)$ to the path integral are
equal but with opposite signs. The contributions of both paths
cancel and the argument can be repeated path by path to show that
the whole path integral vanishes. In a similar way we can 
prove the vanishing  of the kernel element
$K^A_{\hbox{\sevenrm{T}}}(\phi_{+-},\phi_{--})=
K^A_{\hbox{\sevenrm{T}}}(\phi_{-+},\phi_{--})=0$ for
$\phi_{+-}$ and $\phi_{-+}$, because in that case the
corresponding holonomies of  paths and reflected paths differ
by the holonomies of the loops $\alpha^{2n_1+1}_- 
\circle\beta^{2n_2}_+$ and $\alpha^{2n_1}_+ 
\circle\beta^{2n_2+1}_-$, respectively. The relative negative sign
is again the basis for the cancellation of the corresponding
contributions to the path integral.
Notice, however, that the argument cannot prove the vanishing 
of $K^A_{\hbox{\sevenrm{T}}}(\phi_{-+},\phi_{++})$, $K^A_{\hbox{\sevenrm{T}}}(\phi_{+-},\phi_{++})$ or
$K^A_{\hbox{\sevenrm{T}}}(\phi_{+-},\phi_{-+})$. 

\vskip 0.2cm
{\hskip 2.3cm\epsfbox{contodo.ai}}
\vskip 0.1cm
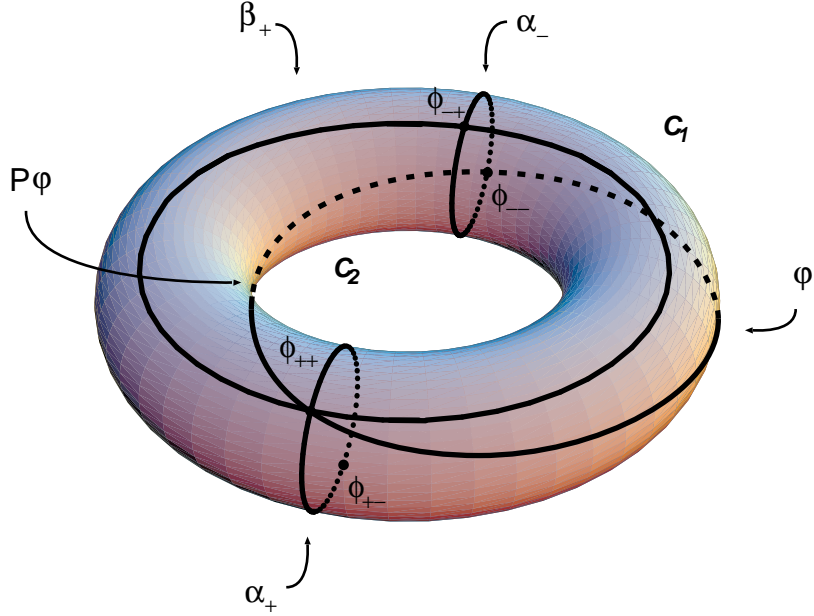
\centerline{\hbox{\rm {\bf Figure 2.} Paths giving oposite contributions
to the path integral kernel}}
\vskip 0.3cm
\end{proof}
Let  $P_{\uparrow\uparrow}$ denote the reflection operator
 with respcet to the crossing points $\phi_{\uparrow\uparrow}$
 where the two circles with  holonomy $iI$ cross each other for $k=1$.

\begin{theorem} Let $k=1$ and $V$  be an  invariant potential
 under reflections $P_{++}$ and $P_{\uparrow\uparrow}$.
  The ground state is unique, $U(P_{++})$ even and has a
  node at $\phi_{++}$.
\end{theorem}
\begin{proof}
In the case $V=0$ the parity behaviour of the vacuum state can 
be obtained from an indirect argument. We know that the ground states 
in absence of potential term 
are holomorphic sections of the line bundle $E_k(T^2,\IC)$
with Chern class $c(E_k)=k$ (see Ref. \cite{3} for a review 
and references therein). Any holomorphic section of $E_k$ can
 have only $k$ single nodes. For $k=1$ only $P_{++}$--even states
can have only one  single  node at $\phi_{--}$, whereas 
$P_{++}$--odd states have at least three-different nodal points. Thus, the
vacuum state is  $P_{++}$--even and has a node at the crossing
of the two circles with holonomy $-I$ (the corresponding 
Abrikosov  lattice has one single vortex). The same state 
is parity odd with respect to $P_{--}$, $P_{+-}$ and $P_{-+}$.
(The property also holds for $k>1$).
Those results can be explicitly checked from the exact
analytic solutions
\be
\begin{array} {ll} 
\displaystyle
\psi_{nl}(\varphi_1,\varphi_2) &={1\over 2\pi}
\left({2\over k}\right)^{1\over 4}
{1\over \sqrt{n!2^n}} {\hbox{\rm e}}^{i{k\over
4\pi}\varphi_1 \varphi_2}\sum_{m\in \IZ} 
{\hbox{\rm e}}^{i{ m}\left(\varphi_1 +2\pi{l\over k}\right)}\\
\displaystyle
&\times H_n\left(\sqrt{k\over 2\pi}\left(\varphi_2+2\pi{m\over
k}\right)\right)
{\hbox{\rm e}}^{-{1\over 4 \pi k}\left(2\pi m+k\varphi_2\right)^2}\\
\displaystyle
& l=0,\cdots,k-1\\
\end{array} 
\ee

However, the  symmetry arguments already introduced in the case of
the planar rotor allows us to  generalize this result for more general
potentials which makes it extremely useful especially for 
non-exactly solvable cases.

The vanishing
of $K^A_{\hbox{\sevenrm{T}}}(\phi_{++},\phi_{--})=0$ for any $T$ implies that
$$\sum_n U(P_{++}) \psi_n(\phi_{++})^\ast
U(P_{++})\psi_n(\phi_{--})=0$$ for any energy level.

If the ground state is degenerate there are at least two states
$\psi_0^+$ and $\psi_0^-$  which are even and odd, respectively,
with respect to $P_{++}$--parity. $\psi_0^+$ vanishes at $\phi_{--}$ 
and $\psi_0^-$ at $\phi_{-+}$, $\phi_{+-}$ and
$\phi_{++}$. We know that the kinetic term contribution
is minimized in a state with a single node at $\phi_{--}$.
Since the potential term is reflection symmetric
it has the same behaviour near the four points. Thus,
the kinetic and potential energies  are
minimized on states with a unique node at $\phi_{--}$ 
instead of three nodes at
$\phi_{-+}$, $\phi_{+-}$ and $\phi_{++}$. From Ritz's variational 
argument  it is obvious that  $\psi_0^+$ and $\psi_0^-$  cannot have the
same energy. The existence of such non-trivial
splitting implies that the vacuum state $\psi_0$ is unique and,
thus,  has a unique node at $\phi_{--}$ and is even with respect
to $P_{++}$--parity and  odd with respect to $P_{-+}$, $P_{+-}$
and $P_{--}$ parities.
\end{proof}

It is also easy to understand this result from perturbation
theory,  because $V$ does not connect
parity even states $\psi_+$ with parity odd states $\psi_-$. In fact,
by parity symmetry 
$<\psi_+|V^n|\psi_->=0$, which implies that there are no corrections
to the parity behaviour of the ground state at any order in perturbation
theory. This result is, thus, compatible with our non-perturbative result
which holds for larger potentials
and tells us that there is no level crossing of ground states
whenever we keep the reflection
symmetry properties of the potentials.

We have found that the vacuum state vanishes at the
intersection of the only pair of circles with holonomy $-I$ . The
singularity of this point explains very explicitly why translation
invariance is completely broken. Only  translations by $2\pi$ can leave
the quantum states and their nodes invariant. For higher values of
$k$ we get similar results but now there are $k$ circles with
holonomy $-I$ in each direction which cross at $k^2$ different
points.  Any of such points is one of the $k$ 
nodes of parity even states under reflections with respect
to the oppossite crossing point of two circles with
unit holonomy. Those nodes give rise to the Abrikosov lattice
of vortices in a type II planar 
superconductor. But now we have different states vanishing at the
different intersections which transform one into each other by the
remaining discrete symmetries. In addition there are linear
combinations of  parity even and odd states for each level which vanish
elsewhere on $T^2$.  Therefore, nothing special happens at that
intersections. Although in  the $k=1$ case the points $\phi_{++}$,
$\phi_{--}$, $\phi_{-+}$
and $\phi_{+-}$ are distinguished points because they are the only
nodes of stationary states, they do not have any physical meaning
because the breaking of translation symmetry is reflected in the fact that
there is a $T^2$ moduli space of $U(1)$--connections $A$ which
generate the same magnetic field but differ by their holonomies, and
for any point $\phi$ of $T^2$ there is a connection whose ground
state vanishes at $\phi$. All these connections and their
corresponding eigenstates are obtained from those of one fixed connection by
translations. Therefore the location of the vacuum nodes in this case
does not have an special meaning. This is in contrast with what
happens in quantum field theories, where the quantum vacua are
really unique and their nodes are very special field configurations
carrying, therefore, a relevant dynamical information \cite{8}--\cite{10}.

The same arguments also apply for higher genus surfaces or surfaces
with holes, giving  very relevant information on the structure of
the Abrikosov lattice for systems with defects.
\vskip.2cm

{{\it Acknowledgements:}} We thank  Luis J. Boya,
 and Fernando Falceto for discussions and collaboration. The work of 
 M.A. is supported by a MEC fellowship (Spain).
We acknowledge CICYT for partial financial support under grant
 AEN97-1680. 

\enddocument
\end